# Investigations on synthesis and characterization of functionalized graphene sheets-polyacrylamide composites

**Ratnesh Pandey, Kalpana Awasthi, R.S.Tiwari and O.N.Srivastava**

DST unit of Nanoscience and Nanotechnology, Department of Physics,
Banaras Hindu University, Varanasi-221005, INDIA

**Abstract**

The functionalized graphene sheets (FGS) - polyacrylamide (PAM) composite films have been prepared by solution cast technique. The FGS have been synthesized by thermal exfoliation of graphite oxide. Several composites with different weight % (between 0 to 25 wt %) of FGS loading in PAM have been prepared and characterized by XRD, SEM, TEM and FTIR spectroscopy. FTIR analysis revealed the existence of the amide linkage (-NH-C=O) between the polymer chain and oxygenated graphene sheets. The electrical and mechanical properties of FGS-PAM composite films have been studied by employing low temperature dc electrical conductivity measurement and nanoindentation technique. A significant increase in both electrical conductivity and mechanical properties (hardness and elastic modulus) is observed at 15 wt% loading of FGS in PAM. As the FGS loading increases from 0 to 25 wt% the conductivity of composite film increases by seven orders of magnitude ($\sim 5.55 \times 10^{-7}$ to 1.42 S/cm). The hardness and elastic modulus of FGS ($\sim$15wt %) PAM film increased four and five fold, respectively as compared to that of PAM film. To our knowledge studies on FGS-PAM composites and also improvement of both the electrical and mechanical properties have not yet been made.

Address correspondence to

hepons@yahoo.com, Ph: +91-542-2368468, Fax: + 91-542-2369889

# 1. Introduction

Graphene, an exotic new nano form of carbon has recently attracted tremendous attention of scientific community due to its unique electrical, mechanical, thermal and optical properties [1]. It is one atom thick monolayer of $sp^2$ hybridized carbon atoms forming six membered rings and is the basic building block of all other graphitic carbon materials [2]. Graphene sheets have extraordinary electrical property, high mechanical strength (>1TPa), high thermal conductivity, (~3000$Wm^{-1}k^{-1}$) and very high surface area (~2600$m^2g^{-1}$). These unique properties make graphene an excellent additive for significant enhancement of electrical mechanical and thermal properties of polymer materials. Graphene sheets have also been used in fabrication of field effect transistors [3], molecular sensors [4], super capacitors [5] and gas storage materials [6]. The first historical isolation of single layer graphene from micromechanical cleavage of highly oriented pyrolytic graphite (HOPG) was reported by Nevoselove et al. [7] in 2004. Currently, several methods have been reported for the preparation of graphene sheets, such as epitaxial growth on SiC wafer [8], chemical vapor deposition of hydrocarbon on metal surface [9], thermal exfoliation of graphite oxide [10] and chemical reduction of graphite oxide [11]etc. Among, them thermal and chemical exfoliation of graphite oxide forms effective ways to prepare graphene in comparatively large quantity and at low cost. Stankovich and co-workers [11] prepared graphene sheets by exfoliation of graphite oxide via ultrasonication treatment, followed by chemical reduction with hydrazine hydrate to modify their transport properties. Recently, graphene sheets have been prepared by arc evaporation of graphite electrodes under hydrogen and argon atmosphere [12] However, obtaining single sheet of graphene in large quantity is still a challenge. The graphite oxide is a layered material produced by oxidation of graphite. Each layer consists of covalently attached oxygen-containing groups such as hydroxyl, epoxy and carboxylic groups [13].

Polymer composites based on graphene are expected to show interesting electrical and mechanical properties. Stankovich et al. [14] reported a general approach for the preparation of graphene-polymer (polystyrene) composites via chemical reduction of isocyanate modified graphite oxide in presence of polymer and studied their electrical properties. At a graphene loading of about 0.15 vol%, the electrical conductivity of composite increases and rapidly rises over a 0.4 vol% range. Tong et al. [15] have prepared the graphene nanosheet / polymer (vinyl chloride o/vinyl acetate) composite using in situ reduction-extractive dispersion technology. When the graphene sheet content is lower than 1.5 vol%, the conductivity of composites was 3-5 order of magnitude higher than that of composite filled with graphite nanosheets from expanded graphite. In another study, Ramanathan et al. [16] have reported that 1 wt% addition of functionalized graphene sheets (FGS) to poly(methyl methacrylate) (PMMA) leads to increase of 80% in elastic modulus and 20% in tensile strength. More recently, Das et al. [17] have studied the mechanical properties of FGS-poly (vinyl alcohol) (PVA) and PMMA composites. A significant increase in both elastic modulus and hardness is observed with the addition of 0.6 wt% of FGS.

In the present investigation, we have prepared FGS- polyacrylamide (PAM) composite films by solution cast technique similar to that followed by us for preparation of CNT-polymer composite film [18]. The nanofiller FGS were synthesized by thermal exfoliation of graphite oxide. Polyacrylamide (PAM) is an insulating polymer. It is easily soluble in water and non polar solvents, having good mechanical strength and air stability. In our study, we have found significant enhancement of electrical and mechanical properties of FGS-PAM composites. The present studies on the preparation of FGS-PAM composites and investigations of their electrical and mechanical properties appear to be first of their type.

## 2. Experimental section

### 2.1 Synthesis of functionalized graphene sheets (FGS)

In the present investigation, FGS were synthesized by thermal exfoliation of graphite oxide. This was prepared using graphite powder according to slightly modified Staudenmaiers method [19]. In a typical experiment, graphite powder (<50µm, 1g) was reacted with strong oxidizing solution of conc. $H_2SO_4$ (18 ml), $HNO_3$ (9ml) (all chemicals from Sigma-Aldrich AR grade) and potassium chlorate (<99%, Merck, 11g) at room temperature for five days under magnetic stirring condition. As obtained graphite oxide solution was washed with distilled water and 10%HCl solution to remove sulphate and other ion impurities. The graphite oxide powder was dried at 80º C under vacuum. As prepared graphite oxide was thermally exfoliated to synthesize FGS by rapid heating under an Ar atmosphere. Graphite oxide powder (200mg) was placed in an alumina boat and inserted in to a 1.5 m long quartz tube with outer diameter of 25mm. The sample was flushed with Ar gas for 15 minutes, and the quartz tube was quickly inserted in to a tube furnace preheated to 1050º C and held in the furnace for 30 seconds. The FGS sample was cooled down to the room temperature under Ar gas flow. As prepared graphite oxide was a brownish powder while its thermally exfoliated version was light weighted shiny black powder.

## 2.2 Preparation of FGS-PAM composites

The FGS-PAM composites were prepared by solution cast technique. The PAM (molecular weight ~$5\times10^6$) was purchased from BDH Chem. Ltd and was used without further processing. The PAM powder (10mg) was dissolved in 25ml of deionized water and stirred for 2h. The FGS was dispersed in distilled water and sonicated for 30 minutes and then added to PAM solution The mixture (FGS-PAM) was maintained under magnetic stirring for 6 hours. The composite of different mass fractions of FGS was prepared by mixing FGS in PAM with different wt%. In the present work, we have mixed 1, 3, 5,10,15,20 and 25wt% of FGS in PAM solution. The resulting FGS-PAM solution was transferred in to polypropylene rectangular boxes and left to dry for two days. The dried FGS-PAM film was peeled off and cut in to size of $1\times1cm^2$ for structural, electrical and mechanical measurements.

## 2.3 Characterization techniques

The structural characterization of as prepared samples of graphite oxide, FGS and FGS-PAM composites, was carried out by powder X-ray diffraction (XRD) recorded on a X'Pert Pro PAN analytical system by using Cu K$\alpha$ with ($\lambda$ =1.5406 Å ) radiation. The microstructures of the samples were examined by scanning electron microscope (SEM) (Philips XL 20) and transmission electron microscope (TEM) (Tecnai 20 $G^2$). For SEM observation, the shiny black material was directly mounted to the sample holder overlaid with dried silver glue which is electrically conductive. Samples for TEM studies were prepared by dispersing a small amount of materials in ethanol with sonication for 10 min. The drops of the dispersion were placed onto a holy carbon grid and dried for 30 minutes. Fourier transform infrared (FTIR) spectra of FGS sample was recorded on small pieces of sample embedded in KBr pellets using Perkin Elmer 100 spectrometer.

## 3. Results and discussion

### 3.1 X ray diffraction analysis

The XRD patterns of graphite, graphite oxide and FGS are shown in figure 1(a). The XRD of graphite shows a sharp peak at 2$\theta$=26.4º. This peak corresponds to (002) reflection of graphite with interlayer spacing ($d_{002}$) 3.41 Å (figure 1(a)). In the typical XRD pattern of graphite oxide shown in figure 1(a), a new peak appears at 2$\theta$=13.2º, corresponding to the

(002) reflection of graphite oxide. The d-spacing increases from 3.41 Å of pristine graphite to 7.42 Å for graphite oxide. The increased spacing indicates the presence of functional groups on the surface of graphite oxide. The XRD pattern of FGS sample (figure 1(a)) shows small broad 002 reflection with interlayer spacing of 3.73 Å, which suggests the formation of few layer graphene. Compared to graphite oxide, the thermally exfoliated graphite oxide shows a decrease in the $d_{002}$ value. This decreased value confirms the decomposition of graphite oxide sheets to graphene sheets. The d-spacing of FGS (3.74 Å), higher than graphite (3.41Å) may be due to presence of residual oxygen containing functional groups. The graphene sheets obtained here are functionalized. In order to obtain the number of layers in graphene sample, we have used Lorentzian curve fitting for (002) reflection of FGS (figure 1(b)). By fitting (002) reflection, we can obtain the average number of layers using Scherer formula [20]. Figure 1(b) shows the Lorentzian fit for the (002) reflection of as prepared FGS sample .The average number of graphene layers in FGS sample is found to be ~five . Figure 2 shows representative XRD of PAM film and FGS-PAM composite film at 15wt% loading of FGS. Polymer film exhibits diffraction peaks at 14º, 17º and 18.5º which correspond to crystalline phase of the polymer. In FGS-PAM nanocomposite sample, the intensity of diffraction peaks of the PAM decreases and peaks become broader. This indicates presence of interactions between polymer chain and FGS. The crystalinity of polymer is expected to decrease which is suggested by the appearance of broad XRD peaks. However, no obvious diffraction peak of graphite oxide or graphite was observed in XRD of FGS-PAM composite film.

**3.2 Microstructural characterization**

The morphology and microstructure of the as prepared FGS and FGS-PAM samples were characterized by SEM and TEM. A representative SEM image of FGS sample is shown in figure 3(a). This shows agglomerated sheets with fluffy morphology. The TEM image of FGS reveals a wrinkled paper like structure (figure 3(b)).These wrinkles on graphene sheets are due to attachment of functional groups on both side of carbon grid and topological defect on graphene sheets. The high resolution TEM (HRTEM) image from edge portion of FGS is shown in figure 3(c). The individual layers of functionalized graphene marked by arrow are clearly described in figure 3(c). The number of graphene layers is measured to be ~ three to five. This is broadly consistent with that obtained from XRD analysis. The inset figure 3(c) is the selected area electron diffraction pattern (SAED) of FGS, showing clear diffraction spot. The diffraction spots were indexed to hexagonal graphite crystal structure. The quality of nanofiller dispersion in the polymer matrix directly correlates with its effectiveness for improving electrical mechanical and other properties. The aspect ratio and loading (wt %) of filler are also intimately interlinked with these properties. The dispersion of FGS in FGS-PAM composite film was observed by SEM. Figure 4 shows typical SEM image of FGS-PAM composite films. For FGS (~5wt %)-PAM composite, it is observed that the dispersion of FGS is not uniform and some agglomerations are observed. A well dispersed distribution of FGS in polymer matrix can be seen in 15 wt% loading of FGS in PAM matrix as brought out by figure 4. In the inset of figure 4 we have shown digital photograph of FGS (~15wt %) -PAM composite film. The TEM image of FGS (~15wt %) - PAM composite elucidated through figure 5. This microstructure shows that FGS are well dispersed in polymer matrix. From SAED pattern of FGS-PAM composite, it is clear that FGS embedded in polymer matrix shows the crystalline nature. The SAED pattern of polymer indicates the crystalline nature of the polymer (inset of figure 5). Thermally exfoliated graphene sheets contain some oxygen containing groups at their edges and on their basal plane. Most of them can be removed during thermal expansion in Ar atmosphere. Moreover some vacancies and topological defects can be simultaneously produced on graphene sheets, because of release of $CO_2$ during thermal expansion. The functional groups and some topological defects on graphene sheet can provide better interaction between C=O and O-H groups with $NH_2$ group present in the long chain polymer.

### 3.3 FTIR analysis

FTIR spectroscopy is one of the sensitive tools for observing the interaction between the FGS and polymer (PAM) through the modifications in the vibrational spectra. A representative FTIR spectrum of FGS is shown in figure 6(a). The observed peak at 3410 $cm^{1}$ corresponds to hydroxyl (-OH) stretching vibration. A small peak at 1730 $cm^{-1}$ is associated with the carbonyl (C=0) stretching of the carboxylic (-COOH) group. The absorption peak appearing at 1560 $cm^{-1}$ in terms of the known results is attributed to the vibration of carbon skeleton of graphene sheet. In addition, the peaks at 1385 and 1126 $cm^{-1}$ are related to C-OH (hydroxyl) and C-O (epoxy), respectively. This observation confirms the presence of oxygen containing functional groups e.g. C=O, C-OH and C-O on the graphene sheets. It is generally recognized that the hydroxyl and epoxy groups are present above and below the basel planes, while the carboxylic groups are bound to the edge of FGS [21]. The oxygen functional groups of FGS can react with the polymer chain and influence the properties of graphene polymer composites. The functionalized graphene sheet is potential material for reinforcement of polymers. FTIR spectra of PAM and FGS (~15 wt %) – PAM composite are shown in figure 6 (b) and (c). In the linear chain polyacrylamide, the functional amide group (-$CONH_2$) is attached to the saturated C-C backbone. For the PAM polymer (figure 6(b)), the absorption bonds at 3370 and 3200 $cm^{-1}$ correspond to the $NH_2$ stretch. The peak at 1645 $cm^{-1}$ is associated to the amide carbonyl (C=O) stretch. The peaks at 2947$cm^{-1}$ (C-H stretching) and 1450-1325$cm^{-1}$ (various CH bending) also confirm the presence of organic groups in PAM polymer chain [22], and small peak at 1193 $cm^{-1}$ is identified as C-N bond stretching. The interaction of FGS with the PAM molecule has been verified by comparing the FTIR spectra of FGS, PAM and FGS (~15 wt %) – PAM composite samples. Interestingly, we observe a peak shift corresponding to amide carbonyl (C=O) stretch (1645 to 1656 $cm^{-1}$) for PAM and FGS-PAM composite. This is presumably due to the interaction between the $NH_2$ bond of PAM and carboxylic group of FGS. The characteristic of – $NH_2$ stretch of PAM is not observed at 3370 and 3200$cm^{-1}$ in polymer composite, which provides further evidence for chemical linkage between FGS and PAM matrix. The peak around 1590 $cm^{-1}$ is associated with the N-H in-plane stretch and the vibration of carbon skeleton of the FGS. A new peak at 1292$cm^{-1}$ corresponds to C-N bond. The peak at 1115 $cm^{-1}$ is due to the C-O stretching of FGS. In passing, it may be pointed out that the above analysis of the FTIR spectra of FGS-PAM is in keeping with other FTIR studies on FGS and on composites involving polymers other than PAM [23, 24]. Thus the above FTIR analysis clearly reveals the chemical cross linking of FGS with PAM chain.

**3.4 Electrical conductivity measurement of FGS-PAM composite**

One of the main challenges of nanocomposites of insulating polymers and conductive fillers is to achieve materials with improved electrical and mechanical properties. The fillers, normally CNTs and graphite nanosheets, exhibited a high aspect ratio which is advantageous for forming conducting networks in a polymer matrix. The dc electrical conductivity of FGS-PAM composites was measured by using four probe technique. The contacts on FGS- PAM films were made by high quality silver paste. The current was supplied from Keithley programmable constant current source (model-220, resolution ~1µA). The room temperature conductivity of PAM polymer film was found to be ~$5.55 \times 10^{-7}$ S/cm. Figure 7 shows room temperature electrical conductivity of FGS-PAM composites as a function of the filler (FGS) loading in wt%. In between 0 to 3wt%, the conductivity changes hardly and remains as ~$10^{-7}$

S/cm. At 5wt%, FGS-PAM composite film conductivity is around $10^5$ S/cm. When FGS content in PAM increases from 5 to 10 wt%, the conductivity of composite increases significantly from $10^{-7}$ to $10^{-2}$ S/cm. A rapid increase in the electrical conductivity of composite material takes place when the conductive filler forms connected network paths through the insulating matrix. The filler wt % content under this transition state is known as percolation threshold. As shown in figure 7, the percolation threshold for electrical conduction of FGS-PAM is observed at around 10 wt % FGS loading in PAM matrix. This is higher than those obtained for polystyrene-graphene [14] and PVA reduced graphite oxide composite [23]. The higher percolation threshold in our case is due to the presence of oxygenated graphene sheets. It is also believed that the PAM coats the graphene sheets and reduces the conduction of FGS-PAM and FGS-FGS contacts, thus a higher concentration is required to reach the percolation threshold. The increase in conductivity with FGS content in PAM indicates that electrical conduction across the composite film is mainly due to graphene network in PAM matrix. The temperature dependent conductivity of PAM and FGS-PAM composite film at different wt% loading of FGS in PAM e.g. 5,10,15,20 and 25wt% have been studied by us from room temperature to 16K. Figure 8 shows a typical plot of conductivity of composite films as a function of temperature. In the case of PAM, the variation of conductivity with temperature indicates insulating behaviour of polymer. At ~5wt% FGS content in PAM, conductivity behaviour is nearly similar to PAM, whereas at 10wt%FGS concentration in PAM the conductivity increases by 5 orders of magnitude. In this case conductivity increases with the increase of temperature showing semiconducting behaviour. The maximum conductivity value has been obtained for FGS (~25wt%)-PAM composite film. The conductivity value for this composite was found to be 1.42 S/cm which is seven orders higher than for PAM film. The electrical conductivity of functionalized graphene sheets was found to be 10.58 S/cm. The increase in conductivity values with respect to temperature suggests the semiconducting nature of FGS as well as its composite. Recently Jun lto et al. [25] have studied the electronic property of oxygen absorbed graphene sheet and they have suggested that the finite energy gap at K point emerges for oxygen absorbed graphene and its value increases with increase of oxygen to carbon (O/C) ratio, resulting in a change of structure from zero to finite band gap semiconductor. In the light of this it is expected that presence of functional groups plays an important role in deciding the nature of conductivity of graphene sheets.

## 3.5 Mechanical property measurements

Mechanical properties such as elastic modules and hardness of PAM and FGS-PAM film have been measured using nanoindentation technique. A Berkovich tip (three-sided pyramidal diamond tip) was used for nano-indentation. Since the mechanical properties extracted from the nano-indentation are sensitive to the tip geometry, the tip area function has to be calibrated before determining the mechanical properties accurately. This was conducted by using a standard quartz sample in the present case, following the standard practice. The nanoindentation experiment was carried out on PAM and FGS (~15wt %)-PAM film. The load vs penetration depth curves of the PAM and FGS (~15wt%)-PAM composites are shown in figure 9. The elastic modulus and hardness of PAM and FGS-PAM composite film were calculated using the Oliver–Pharr method [26] For PAM film the value of elastic modules and hardness was obtained from load vs displacement plot at maximum load of 40mN and was found to be ~ 0.105GPa and ~0.48MPa, respectively. In the case of FGS (~15wt %) PAM composite, the estimated value of elastic modules and hardness are ~0.533GPa and ~ 1.93MPa, respectively. Thus we have obtained fivefold increase in elastic modulus and fourfold increase in hardness of FGS (~15wt %)- PAM composite film as compared to PAM film. It may be suggested that strong interfacial bonding between the matrix and FGS is essential for efficient load transfer to obtain high strength. As it is evident from FTIR result that there is a chemical linkage between the polymer chain (matrix) and FGS which results in the formation of strong interface .The dispersion of FGS in polymer matrix and their interaction (or bonding) are the reasons for the enhancement of mechanical properties of FGS- polymer composite.

## 4. Conclusions

The FGS – PAM composite has been successfully prepared through solution cast technique. The FGS have been synthesized by thermal exfoliation of graphite oxide. The as prepared FGS samples have been characterized by XRD, SEM, TEM and FTIR spectroscopy. The chemical linkage of FGS with PAM matrix has been confirmed by FTIR spectroscopy. The interaction of the FGS and PAM chain is confirmed by the shift of the C=O peak and presence of –NH (at 1592 cm$^{-1}$) and C-N (at 1292 cm$^{-1}$) bonds in the FGS-PAM composite. The present investigation reveals that the electrical conductivity of PAM and FGS (~25wt %) - PAM composite film at room temperature are 5.55x10$^{-7}$ S/cm and 1.42 S/cm, respectively.

This indicates that conductivity of FGS-PAM film increases by seven orders of magnitude as compared to PAM. Good dispersion of FGS in PAM and formation of homogeneous network have been found at 15 wt % loading of FGS in PAM. The increase in conductivity with respect to temperature suggests semiconducting nature of oxygenated graphene sheets as well as its composite with polymer. It has been observed that mechanical properties such as hardness and elastic modulus of the FGS (~ 15 wt %)-PAM composite film have increased by four and fivefold, respectively as compared to the PAM film.

## Acknowledgements

The authors are extremely grateful to Prof.C.N.R. Rao (Chairman, Nano Science Mission, Govt. of India) and Prof. R. Chidambaram (principal scientific advisor Govt.of India) for their encouragements and fruitful discussions. The authors acknowledge with gratitude Dr. A.C. Pandey (Allahabad University) for providing nanoindentation facility. The financial support from DST: UNANST, Council of scientific and industrial research (CSIR), University grant commission and Ministry of New and renewable energy (MNRE), New Delhi, India is gratefully acknowledged.

**Figure captions**

**Figure 1.** (a) Representative XRD of graphite, graphite oxide, functionalized graphene sheets. (b) Single Lorentzian fit of FGS for 002 reflection.

**Figure 2.** Typical XRD of PAM and FGS (~15wt %) - PAM composite films.

**Figure 3.** Microstructural characterization of FGS: (a) Scanning electron micrograph of FGS. (b) Transmission electron micrograph of FGS. (c) High resolution transmission electron image of FGS, inset shows SAED pattern of FGS.

**Figure 4.** Representative scanning electron micrograph of FGS (~15wt %) PAM composite, inset shows digital photograph of FGS-PAM composite film.

**Figure 5.** Typical TEM micrograph of FGS (~15wt %) -PAM composite.

**Figure 6.** FTIR spectra of (a) FGS, (b) PAM polymer film and (c) FGS (~15wt %)-PAM composite film.

**Figure 7.** Variation of electrical conductivity of FGS-PAM composite film with FGS loading.

**Figure 8.** Conductivity vs temperature plot of FGS and FGS-PAM composite films.

**Figure 9.** Load–penetration depth curves of PAM and FGS (~15 wt %)-PAM composite films.

**References**

[1] Geim A K and Novoselov K S. 2007 *Nat. Mater.* **6** 183

[2] Katsnelson M I 2007 *Mater. Today* **10** 1

[3] Wu Y Q, Ye P D, Capano M A, Xuan Y Sui, Qi Y M, Cooper J, Shen A, Pandey T, Reifenberger D and Prakash G R 2008 *Appl. Phys. Lett.* **92** 092102

[4] Robinson J T, Perkins F K, Snow E S, Wei Z and Sheehan P E 2008 *Nano Lett.* **8** 3137

[5] Wang Y, Shi Z, Huang Y, Ma Y, Wang C, Chen M and Chen Y 2009 *J. Phys. Chem.C* **113** 13103

[6] Schedin F, Geim A K, Morozov S V, Hill E W, Blake P, Katsnelson M I and Novoselov K S 2007 *Nat. Mater.* **6** 652

[7] Novoselov K S, Geim A K, Morozov S V, Jiang D, Zhang Y, Dubonos S V, Grigorieva I Vand Firsov A A 2004 *Science* **306** 666

[8] Juang Z Y, Wua C Y, Lo C W, Chen W F, Huang Y C, Hwang J C, Chen F R, Leou K C and Tsai C H 2009 *Carbon* **47** 2026

[9] Yuan G D, Zhang W J, Yang Y, Tang Y B, Li Y Q, Wang J X, Meng W, X M, He Z B, C Wu M L, Bello I, Lee C S and Lee S T 2009 *Chem. Phys. Lett.* **467** 361

[10] McAllister M J, LiO J L, Adamson D H, Schniepp H C, Abdala A A, Liu J, Herrera-Alonso M, Milius D L, Car O R and Prud'homme R K 2007 *Chem. Mater.* **19** 4396

[11] Stankovich S, Dikin D A, Piner R D, Kohlhaas K A, Kleinhammes A, Jia Y Y, Wu Y, Nguyen S T, Ruoff R S, 2007 *Carbon* **45** 1558

[12] Subrahmanyam K S, Panchakarla L S, Govindaraj A and Rao C N R 2009 *J. Phys.Chem.* C **113** 4257

[13] Lerf A, He H, Forster M and Klinowski J . 1998 *J. Phys. Chem. B* **102** 4477

[14] Stankovich S, Dikin D A, Dommett G H B, Kohlhaas K M, Zimney E J, Stach E A, Piner R D, Nguyen S T and Ruoff R S 2006 *Nature* **442** 282

[15] Tong W, Luo G, Fan Z, Zheng C, Yan J, Yao C, Li W, and Zhang C 2009 *Carbon* **47** 2290

[16] Ramanathan T, Abdala A A, Stankovich S, Dikin D A, Herrera-Alonso M, Piner R D, Adamson D H, Schniepp H C, Chen X, Ruoff R S, Nguyen S T, Aksay I A, Prud'homme R K, and Brinson L C 2008 *Nat. Nanotechnol.* **3** 327

[17] Das B, Prasad K E, Ramamurty U and Rao C N R 2009 *Nanotechnology* **20** 125705

[18] Awasthi K, Awasthi S, Srivastava A, Kamalakaran R, Talapatra S, Ajayan P M and Srivastava O N 2006 *Nanotechnology* **17** 5417

[19] Staudenmaier L Verfahren zur Darstellung Der Graphitsäure 1898 *Ber. Dtsch. Chem. Ges.* **31** 1481

[20] Subrahmanyam K S, Vivekchand S R C, Govindaraj A and Rao C N R 2008 *J. Mater. Chem.* **18** 1517

[21] Schniepp H C, Li L J, Mcallister M J, Sai H, Herrera- Alonso M, Adamson D H, Prud'homme R K, Car R, Saville D A and Aksaya L A 2006 *J. Phys. Chem. B* **110** 8535

[22] Solpan D, Torun M and Guven O 2008 *J. Appl. Poly. Sci*. **10** 3787

[23] Salavagione H J, Martinez G and Gomez M A 2009 *J. Mater. Chem.* **19** 5027

[24] Park S, Dikin D A, Nguyen S Tand Ruoff S 2009 *J. Phys. Chem. C* **113** 15801

[25] Ito J, Nakamura J and Natori A 2008 *J. Appl. Phys.* **103** 113712

[26] Oliver W C and Pharr G M 1992 *J. Mater. Res.* **7** 1564

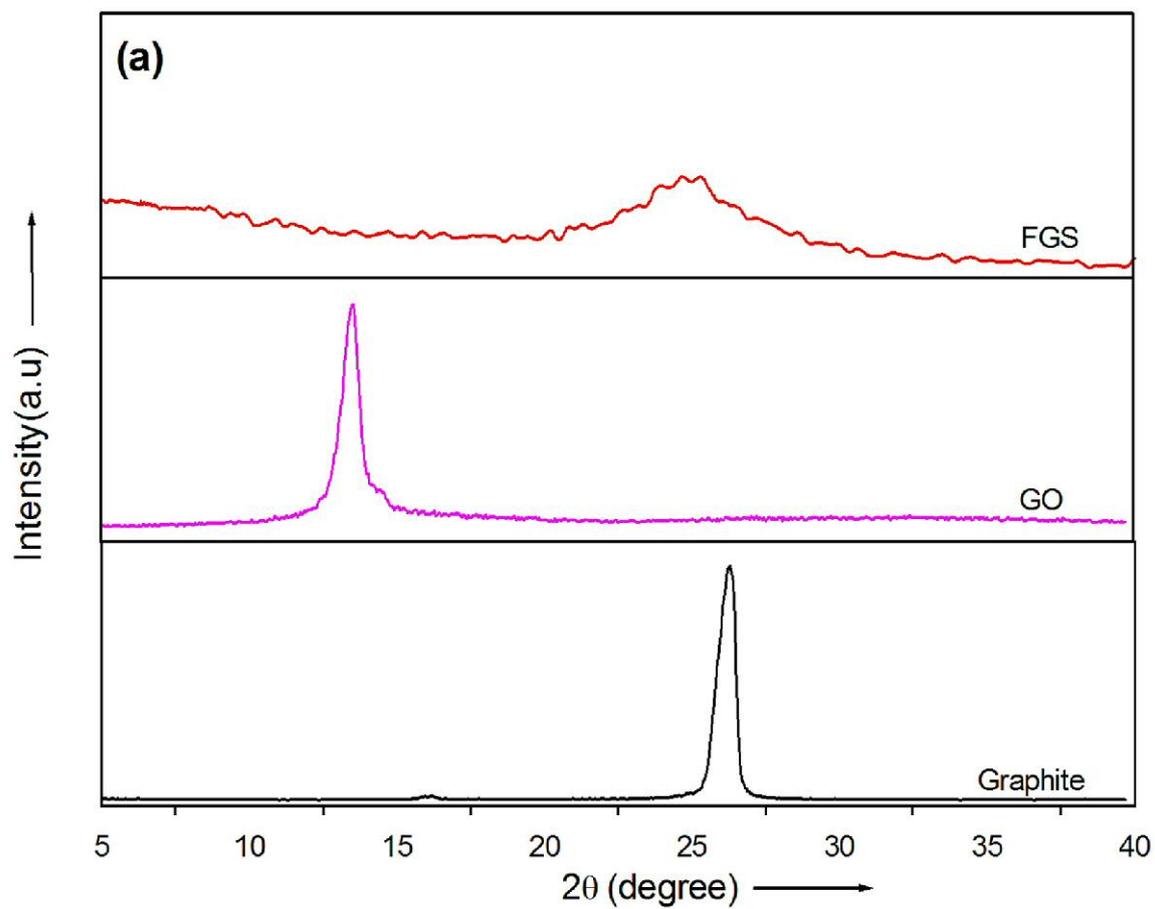

**Figure 1a (Figure 1a.tif)**

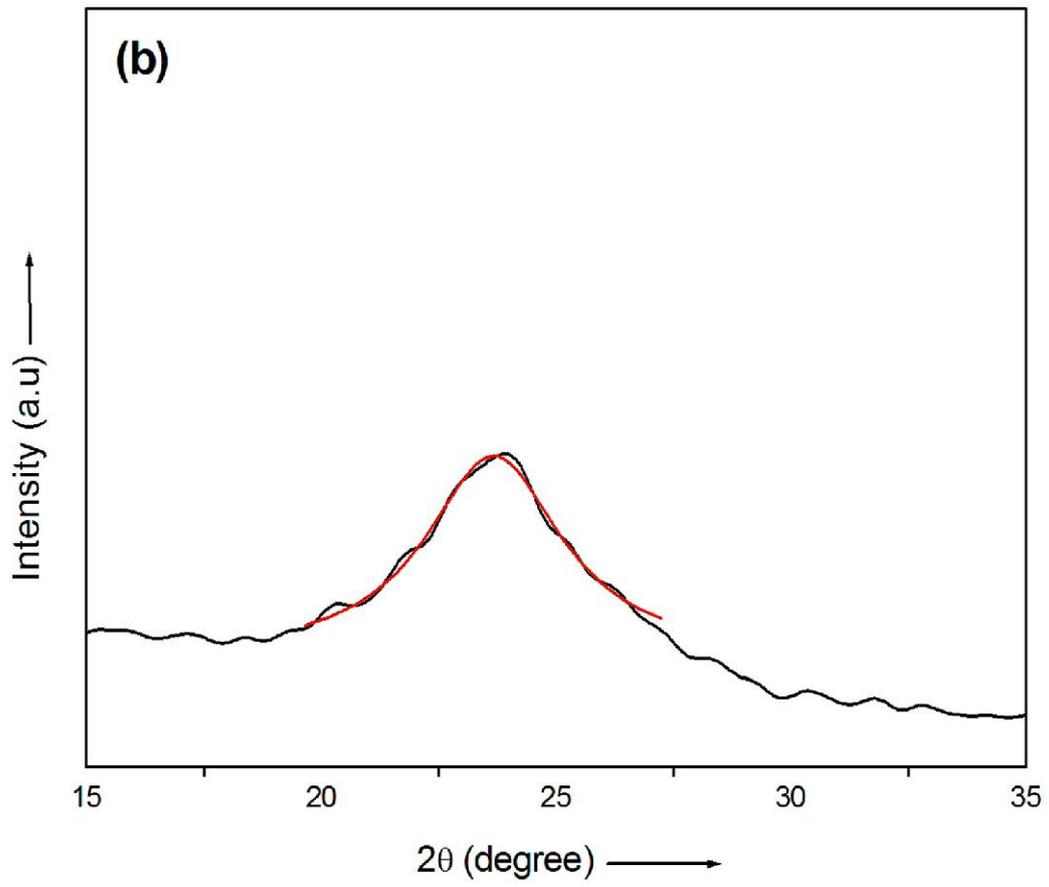

**Figure 1b (Figure 1b.tif)**

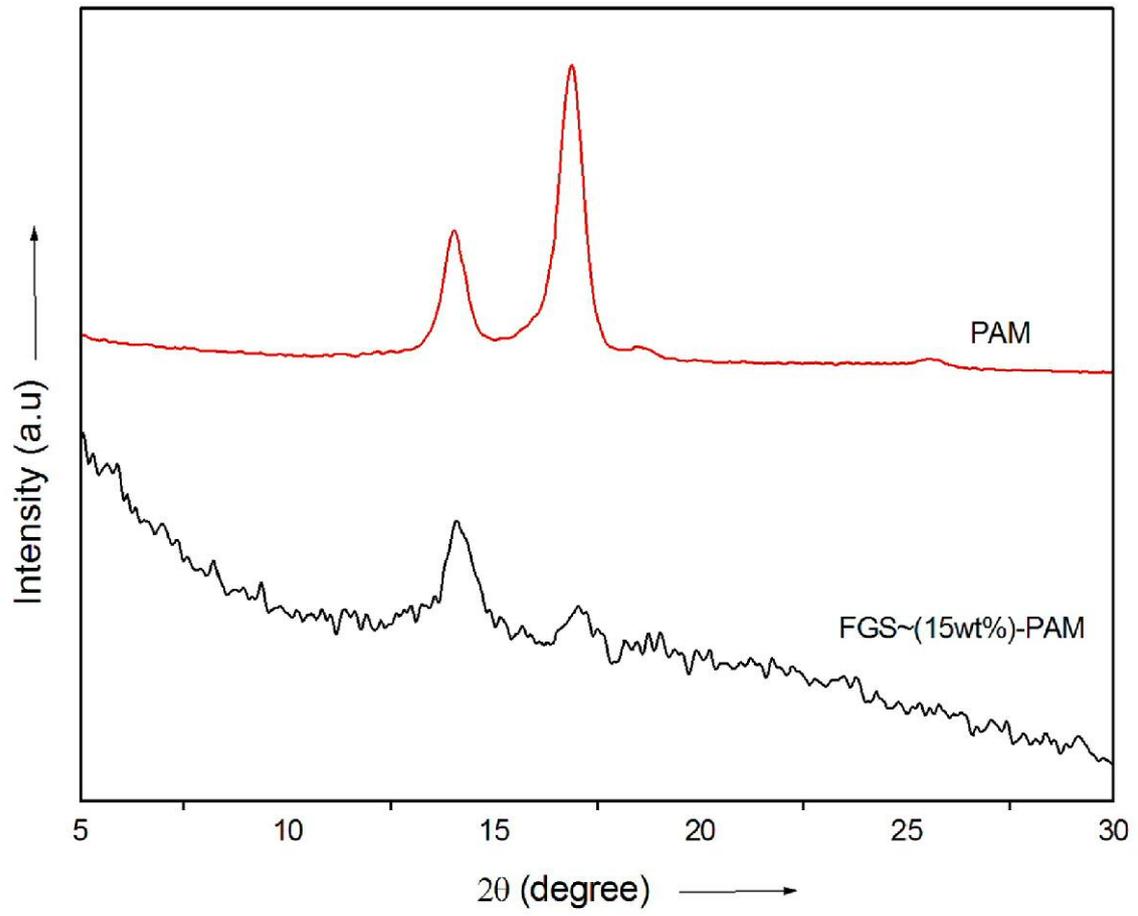

**Figure 2 (Figure. 2.tif)**

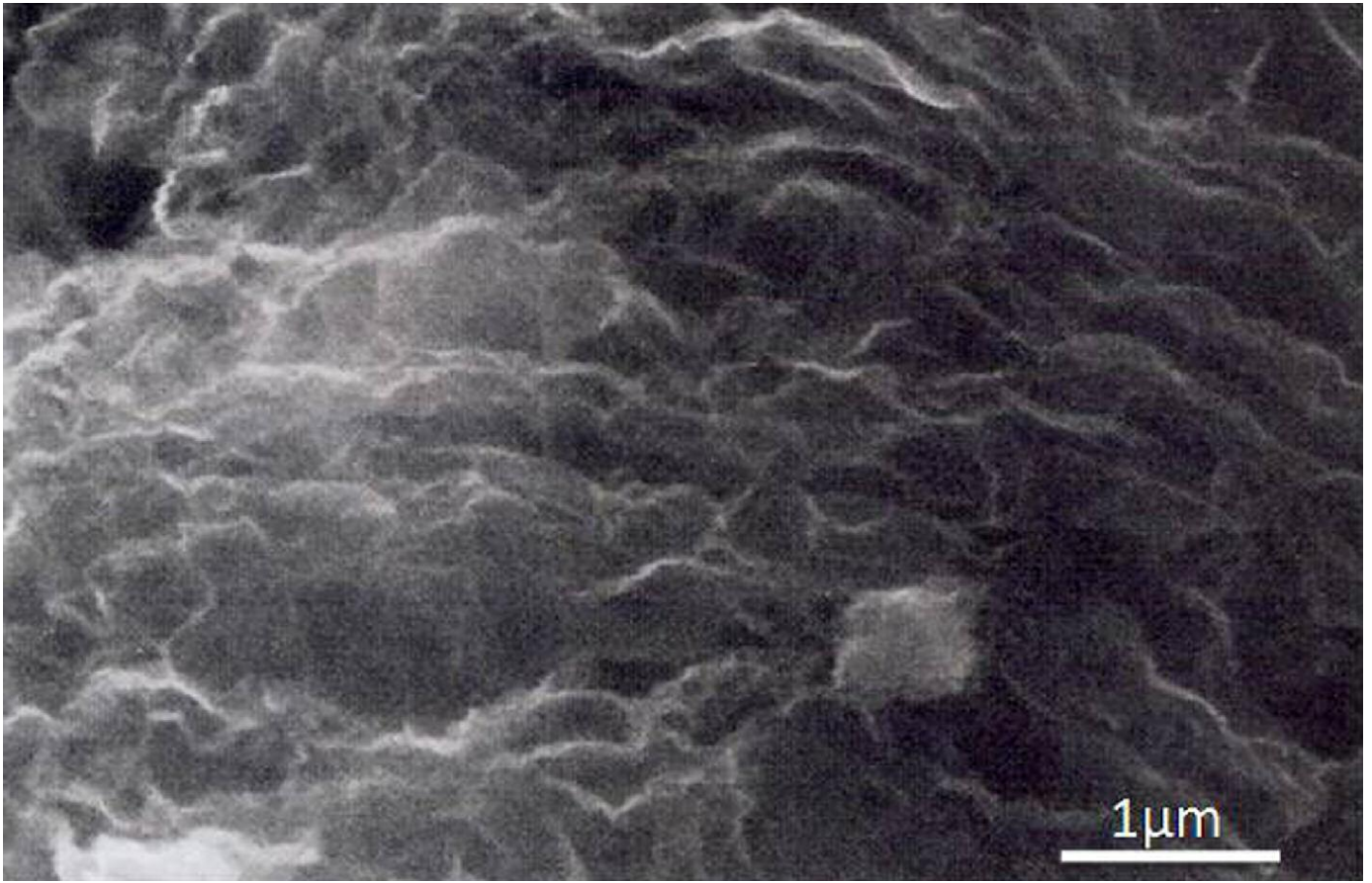

**Figure 3a (Figure 3a.tif)**

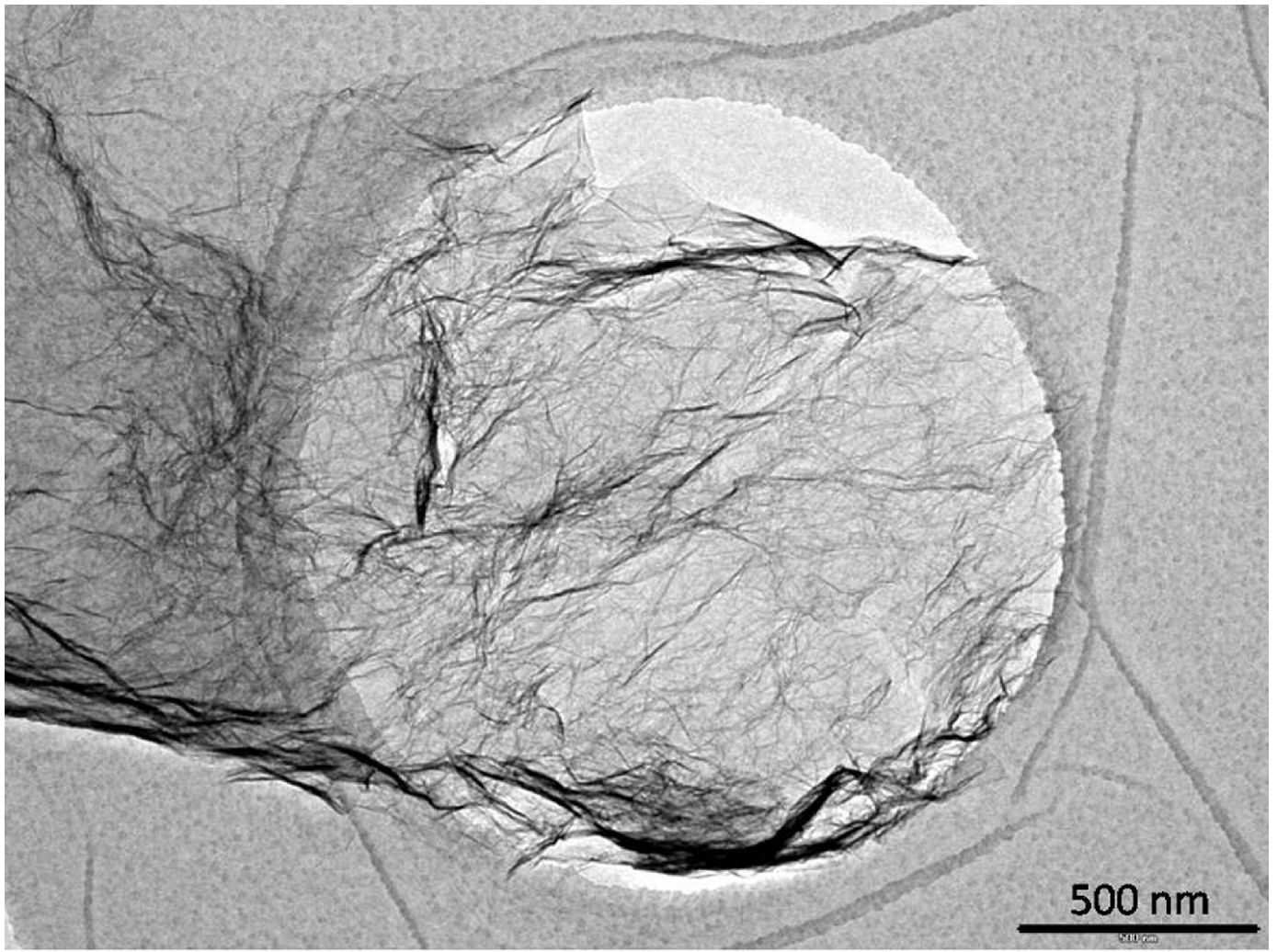

**Figure 3b (Figure 3b.tif)**

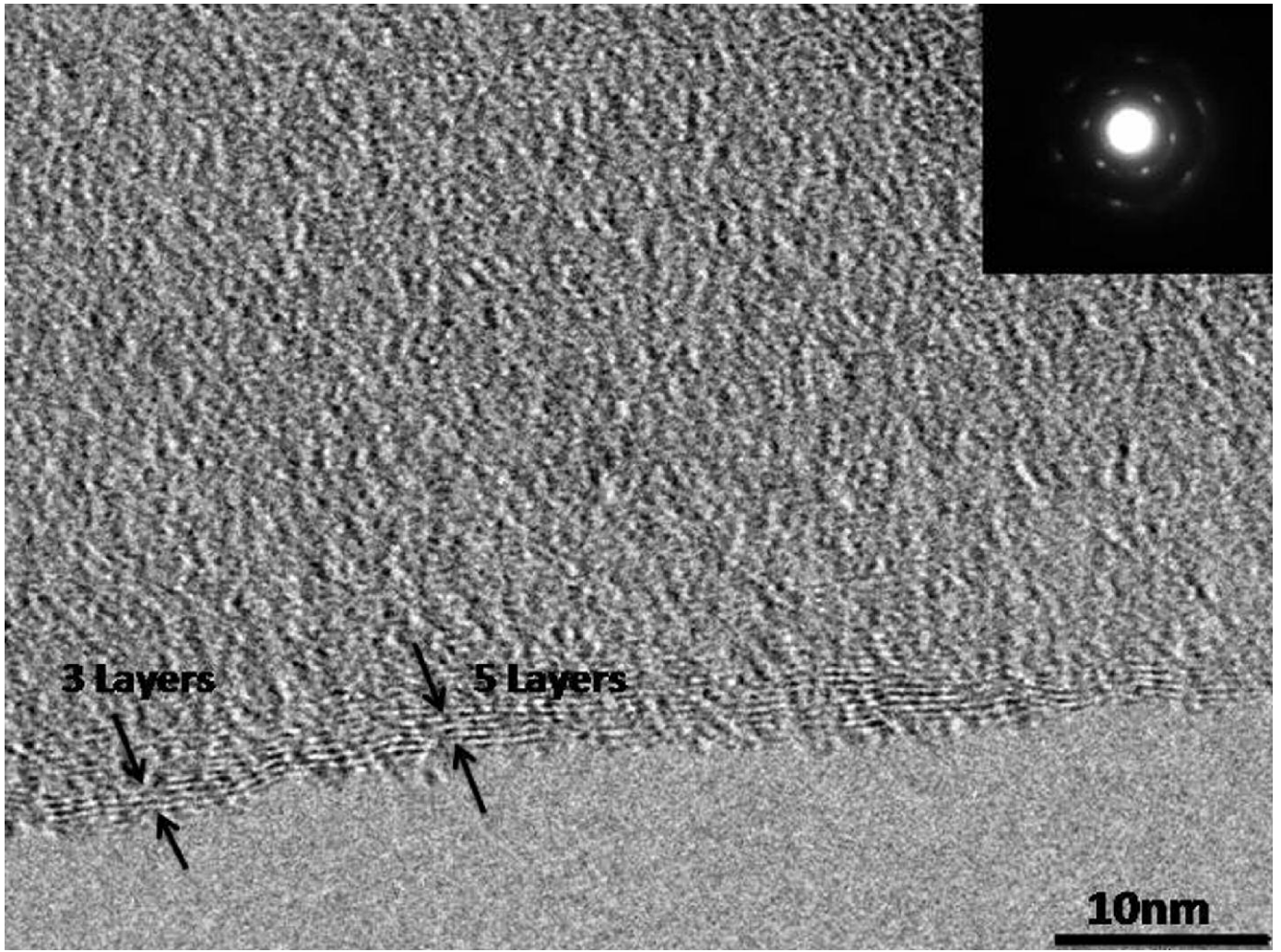

**Figure 3c (Figure 3c.tif)**

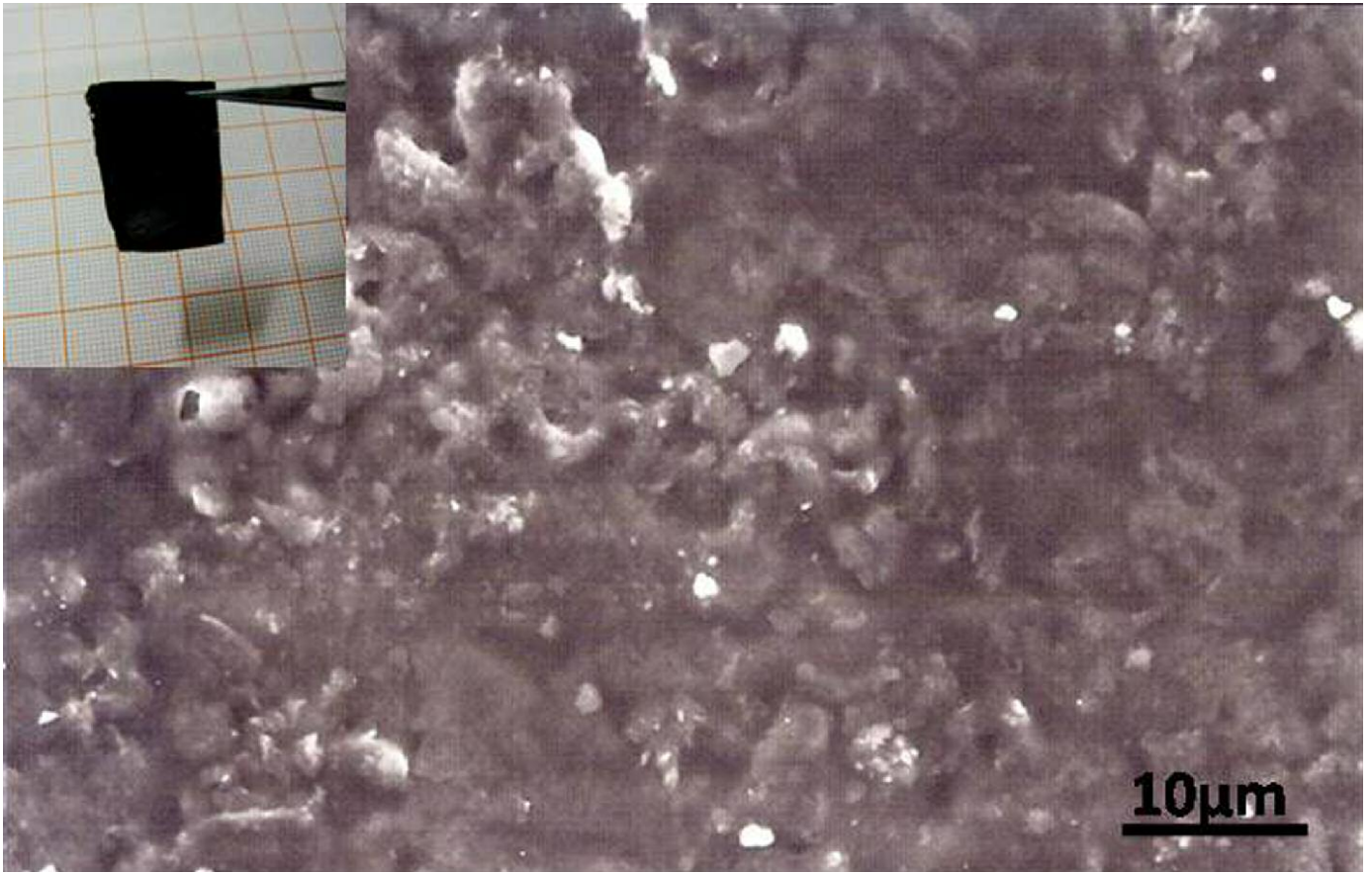

**Figure 4 (Figure 4..tif)**

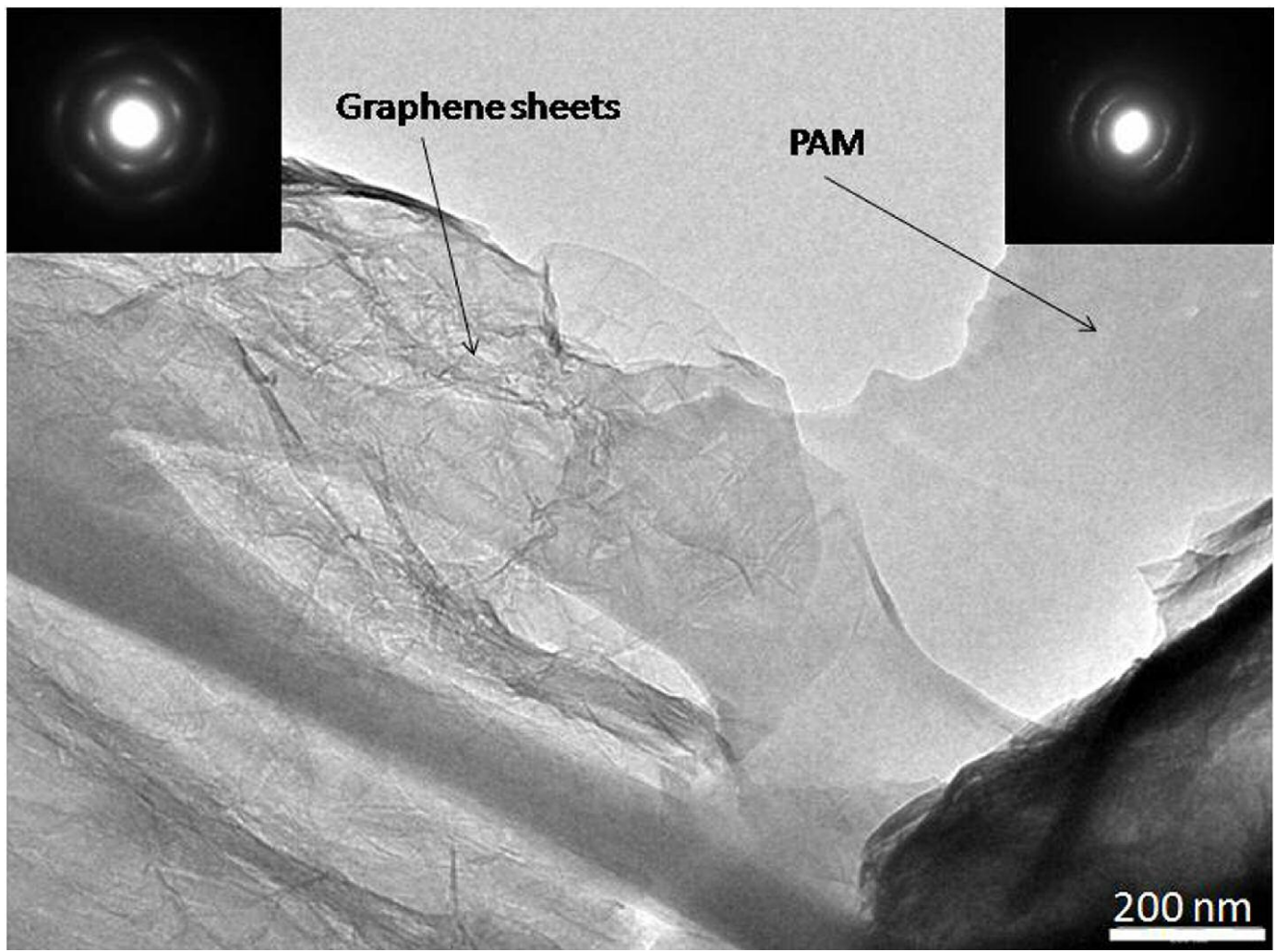

**Figure 5 (Figure 5.tif)**

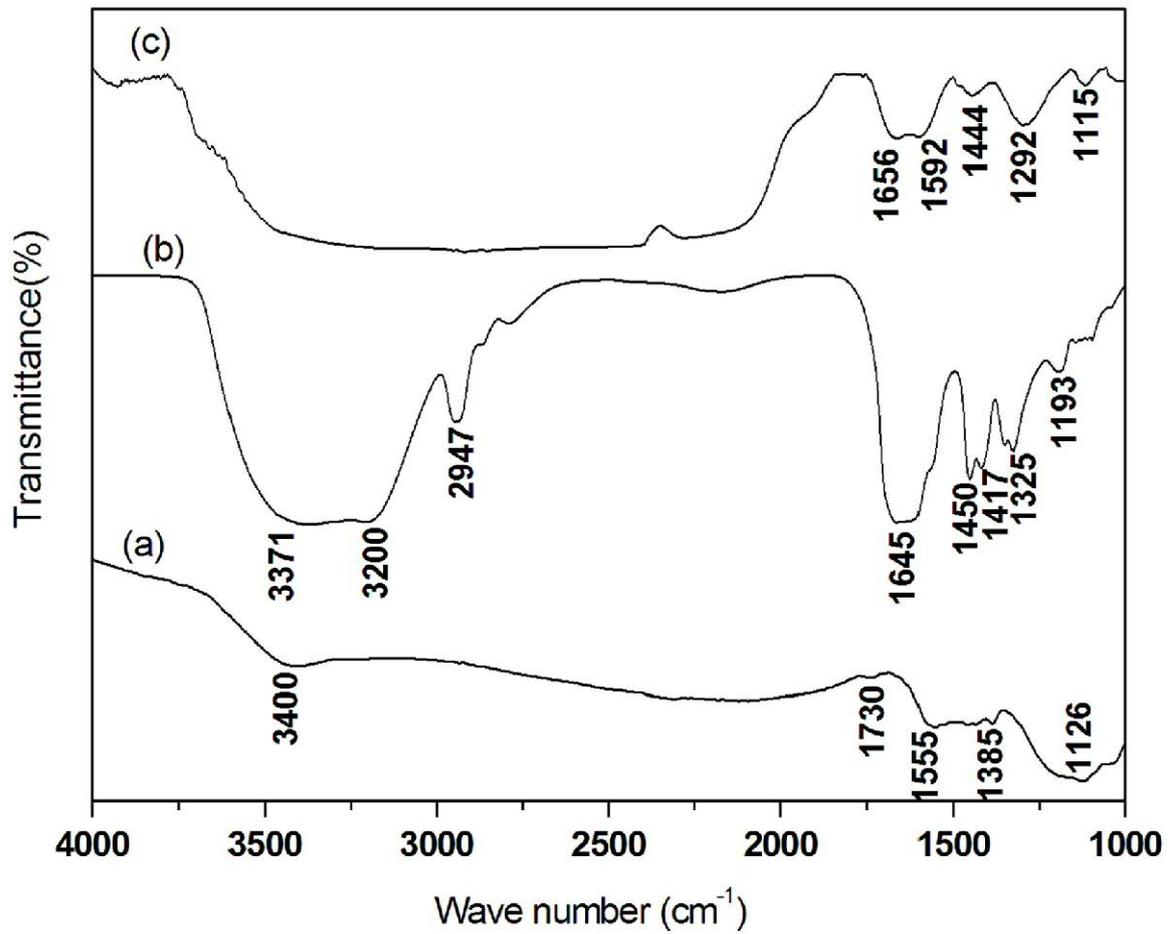

**Figure 6 (Figure.6.tif)**

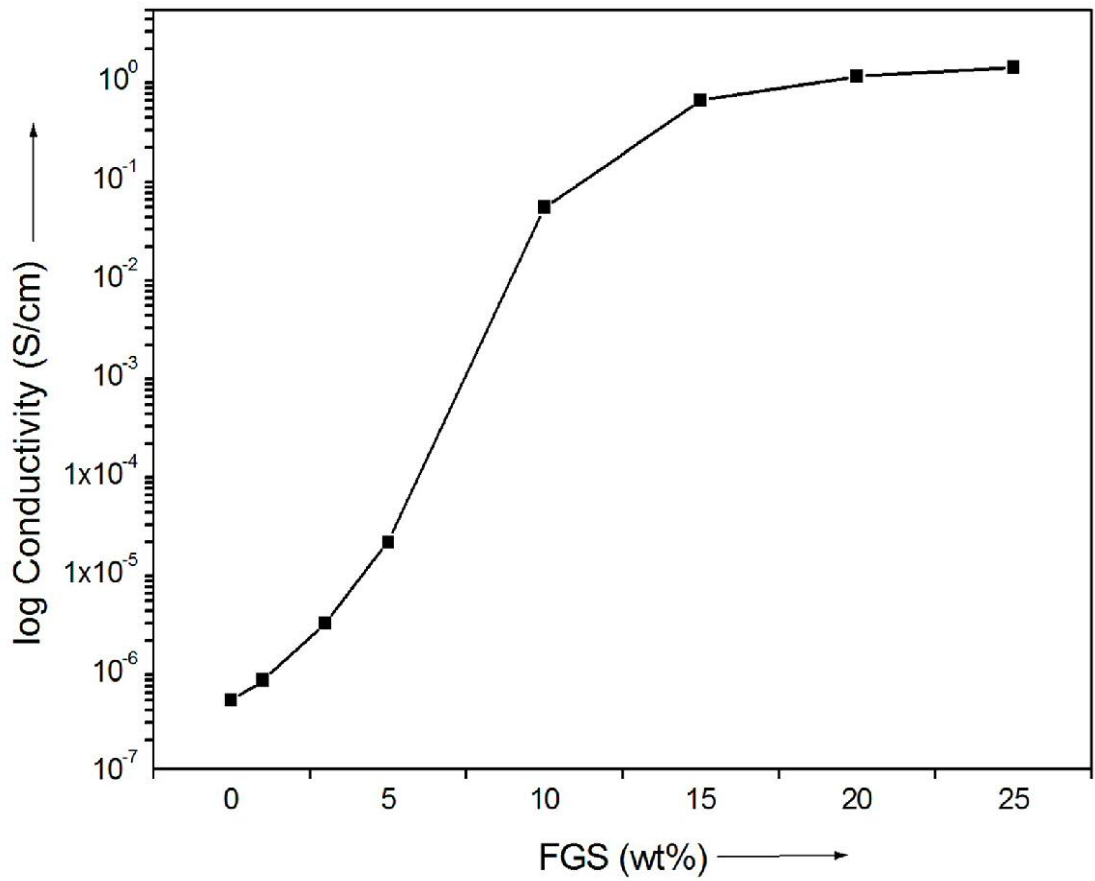

**Figure 7 (Figure 7.tif)**

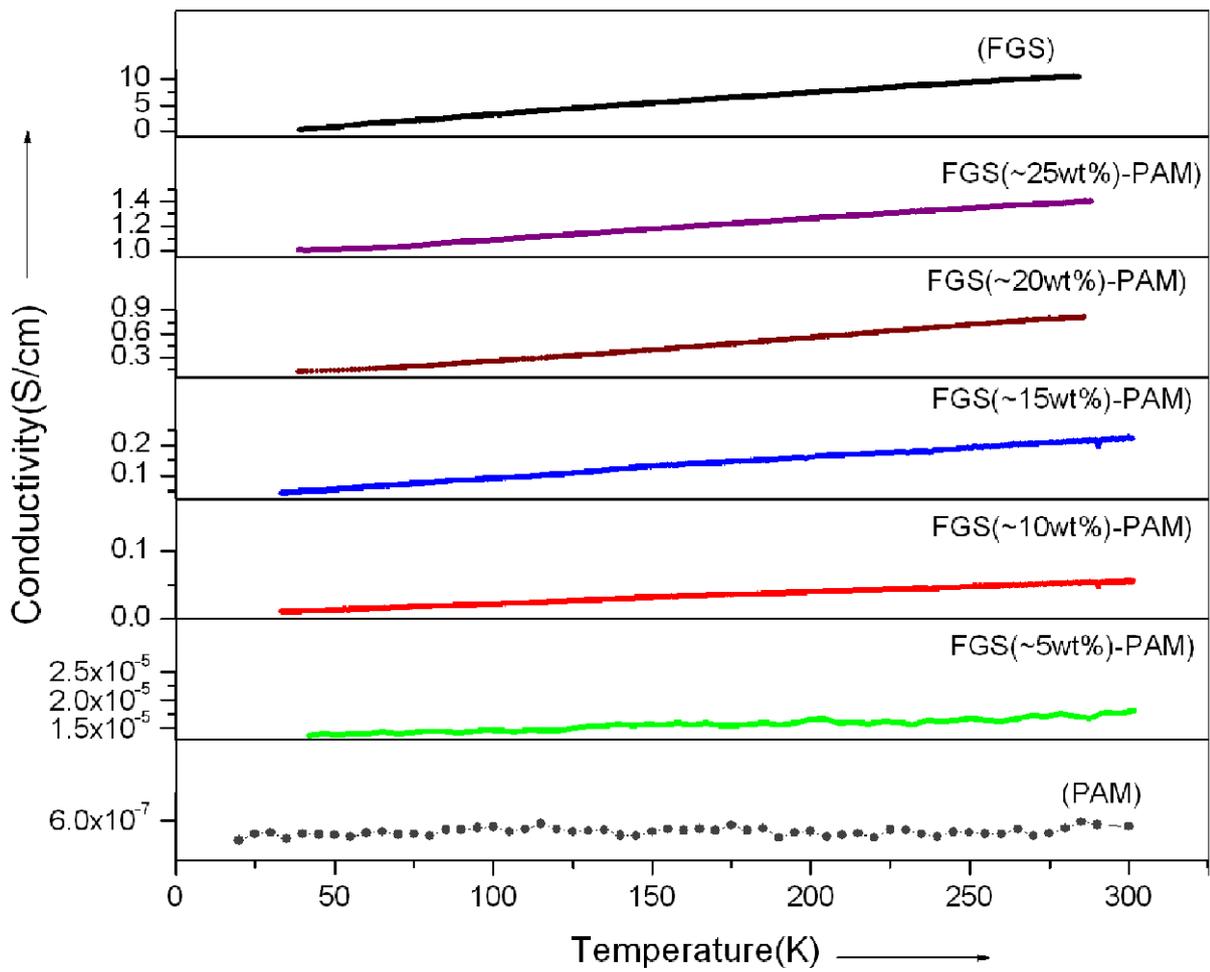

**Figure 8 (Figure 8.tif)**

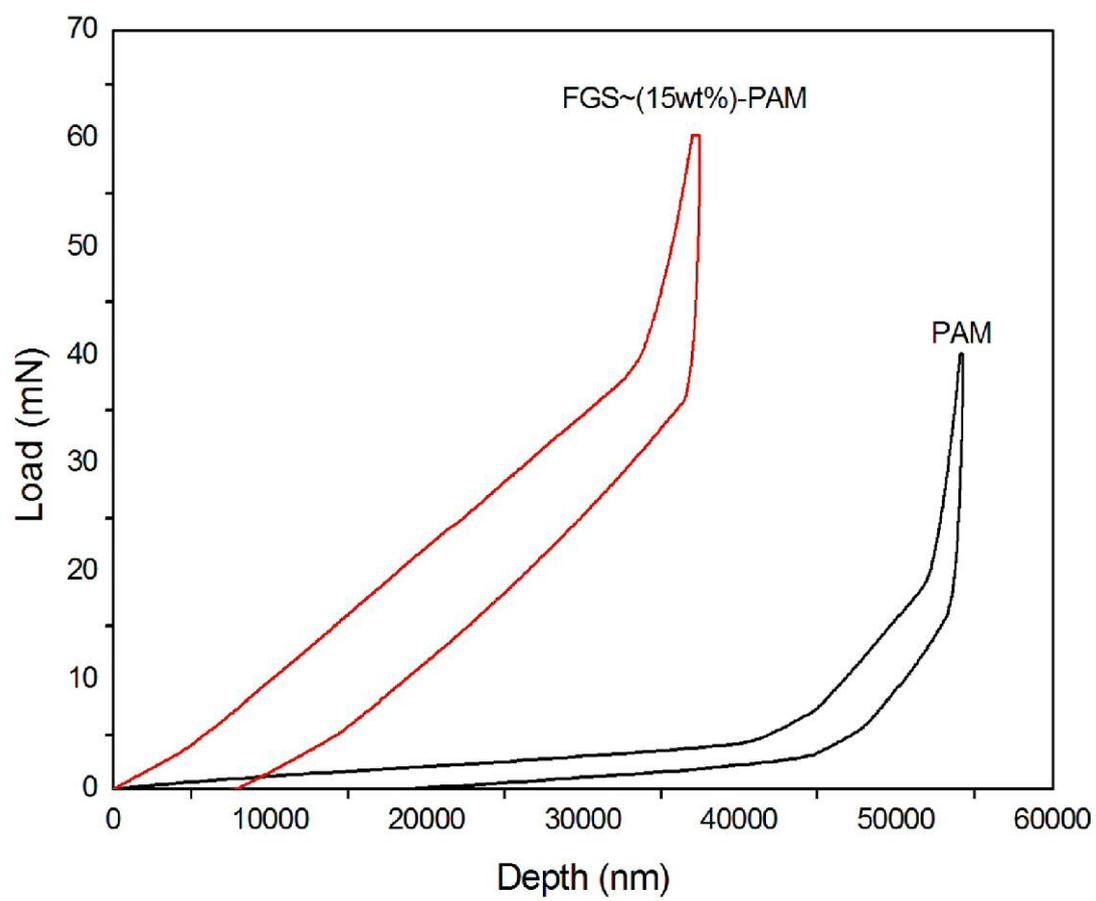

**Figure 9 (Figure. 9.tif)**